\documentclass[pra,amsmath,amssymb, one column]{revtex4}
\usepackage{mathrsfs}
\usepackage{amssymb}

\input epsf.tex
\usepackage{graphicx}
\usepackage{amsthm}

\begin{document}

\title{Randomness determines practical security of BB84 quantum key distribution}
\author{Hong-Wei Li$^{1,2,3}$, Zhen-Qiang Yin$^{1,3,*}$,  Shuang Wang$^{1,3}$, Yong-Jun Qian$^{1,3}$,  Wei Chen$^{1,3}$,  Guang-Can Guo$^{1,3}$,  Zheng-Fu Han$^{1,3,**}$ }

 \affiliation
 {$^1$ Key Laboratory of Quantum Information, University of Science and Technology of China, Hefei, 230026,
 China\\
  $^2$ Zhengzhou Information Science and Technology Institute, Zhengzhou, 450004,
  China\\
 $^3$Synergetic Innovation Center of Quantum Information $\&$ Quantum Physics, University of Science and Technology of China, Hefei, Anhui 230026}

 \date{\today}
\begin{abstract}

Unconditional security of the BB84 quantum key distribution protocol
has been proved by exploiting the fundamental laws of quantum
mechanics, but the practical quantum key distribution system maybe
hacked by considering the imperfect state preparation and
measurement respectively. Until now, different attacking schemes
have been proposed by utilizing imperfect devices, but the general
security analysis model against all of the practical attacking
schemes has not been proposed. Here, we demonstrate that the general
practical attacking schemes can be divided into the Trojan horse
attack, strong randomness attack and weak randomness attack
respectively. We prove security of BB84 protocol under randomness
attacking models, and these results can be applied to guarantee the
security of the practical quantum key distribution system.

\end{abstract}
\maketitle

  Quantum key distribution (QKD)
\cite{reviewBB84} is the art of sharing secret keys between two
remote parties Alice and Bob, unconditional security
of which is based on the fundamental laws of quantum mechanics. The detailed
security analysis has been proved by applying the entanglement
distillation and purification (EDP) technology \cite{edp1,edp2} and the von Neumann
entropy theory \cite{re1,re2,re3} respectively. However, unconditional security of the QKD protocol
has an important assumption, which requires Alice and Bob have random numbers to control the classical bit encoding and measurement bases selection, and it can be easily proved
that security of the final key can't be guaranteed if input random numbers are controlled or known by the eavesdropper Eve.
In recent years, practical QKD system was
attacked by considering the imperfect state preparation and
measurement respectively \cite{practical}. More generally, the
practical attacking scheme can be divided into three different
types. The first type
is considering the Trojan horse attack
\cite{trojan},  where the signal state combining with the trojan horse state can be assumed to be high dimensional state modulation. Note that Eve can measure one dimension of the modulated high
dimensional state to get all of the secret key information without
being discovered, thus Alice and Bob should apply the dimension
filter (such as wavelength filter) to avoid this attack.

The second type is the strong randomness attack, which
considers some of the input random numbers are totally controlled
by the eavesdropper Eve. Such as the multi photon state can be
attacked by applying the photon number splitting (PNS) attack \cite{PNS1,PNS2}, where the multi photon encoding quantum states can be assumed to be known by
Eve. Another example is the detector blinding attack
\cite{blind1,blind2}, where Eve can easily mount the man-in-the-middle (MITM) attack by converting the avalanche photodiodes (APDs) into linear
mode. The detectors
have the count iff Bob's bases selection is equal to Eve, which
means that the bases selection in Bob's side are controlled by
Eve. Recently we propose the probabilistic blinding attack model  \cite{probability},
where Eve partly applies the blinding attack to avoid being catched by
detecting the current parameter. In the strong randomness attack
model, the final secret key rate should delete all of the counting
events known by Eve. The GLLP
\cite{gllp}
secret key rate and the decoy state method \cite{decoy1,decoy2,decoy3} can be assumed to delete all of the multi photon pulse counting
result, and only the single photon counting event can generate the
final secret key. While the probabilistic blinding attack can be assumed to delete all
of the blinding counting results, and only the non-blinding counting
event can generate the final secret key. In the strong randomness
attack model, the previous secret key rate \cite{winter} formula can
be modified to
\begin{equation}
\begin{array}{lll}
R\geq pS(a|E)-fh(e),
\end{array}
\end{equation}
where $p$ is the probability of valid counting result, which can't
be controlled by Eve. $a$ is Alice's measurement outcomes, $E$ is
Eve's auxiliary quantum system, $S(a|E)=S(a,E)-S(E)$ is the
conditional von Neumann entropy, which demonstrates Eve's
uncertainty about Alice's key bit $a$. $e$ is the practical bit
error rate, $h(e)=-elog_2e-(1-e)log_2(1-e)$ is the classical Shannon
entropy function, $f\geq1$ is the error correction efficiency. If we
consider the PNS attack, $p$ can be illustrated by the single photon
counting rate, $S(a|E)$ can be estimated by the single photon error
rate.

The third type is the weak randomness attack, which considers input
random numbers are partly controlled by Eve \cite{weak}. Such as the wavelength dependence about the beam splitter will introduce the wavelength
attack \cite{wavelength}, where Eve can apply different wavelengths to control Bob's
bases selection. Since the practical beam splitter and different wavelengths may
only have partial correlation, which means the beam splitter coupling ratio can't reach $0$ and
$1$ with two different bases, thus Eve can only partly control bases
selection. Another example is the time shift attack \cite{time}, where
Eve controls the APDs detection efficiency by controlling the photon arriving time, thus Eve has the advantage to guess the measurement outcomes.
Since the
practical time shift attack will introduce nonzero error rate, the classical bit
encoding can be assumed to be partly known by Eve correspondingly.

Now, the Trojan horse attack can be avoided by
applying the dimension filter before the state modulation and measurement, which can be used to
prevent Eve's Trojan horse light. The strong randomness attacking model has
also been analyzed by applying the strict post processing, where we only need to precisely estimate $p$ and $S(a|E)$.
However, the weak randomness
attacking model has not been analyzed until now.
In this work, we prove security of the practical QKD system with
weak
input random numbers, which can affect the classical bit encoding
and bases selection respectively. We give two security analysis
models, the first is based on one post processing step, where all of
the measurement outcomes should be applied one time error
correction and privacy amplification.
While the
second is considering two post processing steps, where the measurement outcomes in two bases should be applied post processing respectively.
If we only consider bit encoding weak
randomness, two different methods can get the same secret key rate.
But, if we consider the bases selection weak randomness, the analysis result show that two post
processing steps can generate much more secret key.
Our analysis models can be applied in
several attacking schemes, such as the wavelength attack and the
time shift attack. Combining with the previous three attacking models, security of the practical QKD system can be
evaluated completely. Thus, our analysis result can be applied to
estimate security of the practical QKD system, which can be employed
to build the practical QKD system security standardization.

\textbf{BB84 QKD protocol with weak randomness -} In the BB84
protocol, there are two binary input bits $x_1$ and $x_0$ in Alice's
side, which can be used to select the state preparation bases and
encoding classical bits respectively. While the state measurement
side Bob needs one binary input bit $y$ to select the measurement
basis. After the quantum state preparation and measurement, Alice
and Bob should apply the bases sifting process to save the same
bases case ($x_1=y$). Thus, in the security analysis model, the
input randomness can be divided into two sets, the first set can be
used to decide the encoding classical bit selection $x_0$, while the
second set can be used to decide the encoding and decoding bases
selection $x_1$ (or $y$). Since Alice and Bob should publicly
compare $x_1$ and $y$ to save the same value, we can only consider
Eve has partial knowledge about the bases selection $x_1$ before the
state measurement, the security analysis model can be simplified
correspondingly. Thus we can only assume weak random numbers $x_0$
and $x_1$ to control the encoding classical bit and bases selection
respectively, the detailed analysis model is given in Fig. 1.

\begin{figure}[!h]\center
\resizebox{9cm}{!}{
\includegraphics{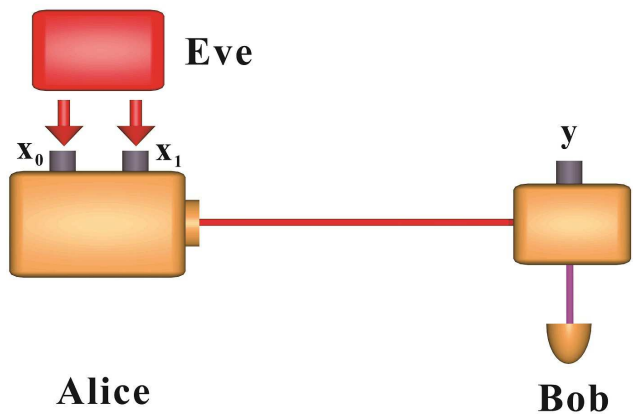}}
\caption{Weak randomness QKD model, where $x_0$ decides the encoding
classical bit, $x_1$ decides the encoding bases selection, $y$
decides the measurement bases selection. In the weak randomness QKD
model, Eve has the advantage to guess the classical bit encoding
$x_0$ and the basis selection $x_1$. }
\end{figure}

In the weak randomness model, the weak random numbers $x_0$ and
$x_1$ can be controlled by two different sets of hidden variables
$\lambda_{0}$ and $\lambda_{1}$ as the following equations,

\begin{equation}
\begin{array}{lll}
p(x_0)=\sum_ip_{\lambda_0=i}p(x_0|\lambda_{0}=i),\\
p(x_1)=\sum_jp_{\lambda_1=j}p(x_1|\lambda_{1}=j),
\end{array}
\end{equation}
where $\lambda_{0}$ and $\lambda_{1}$  are hidden variables
controlled by Eve, $p(x_0=0)$ is the probability that Alice encodes
classical bit $0$, while $p(x_0=1)=1-p(x_0=0)$ is the probability
that Alice encodes classical bit $1$. Similarly, $p(x_1=0)$ is the
probability that Alice applies the rectilinear
 encoding basis,  $p(x_1=1)=1-p(x_1=0)$ is the probability that
Alice applies the diagonal
 encoding basis. Note that two sets of hidden variables $\lambda_{0}$ and $\lambda_{1}$ should satisfy $\sum_ip_{\lambda_0=i}=\sum_jp_{\lambda_1=j}=1$.
However, even if the practical experimental realization can observe
$p(x_0)=\dfrac{1}{2}$ and $p(x_1)=\dfrac{1}{2}$ respectively, we
still can't guarantee
$p(x_0|\lambda_{0}=i)=p(x_1|\lambda_{1}=j)=\dfrac{1}{2}$ for
arbitrary hidden variables $\lambda_{0}=i$ and $\lambda_{1}=j$.
Thus, the the aforementioned security analysis model based on
perfect random input numbers can't be satisfied directly, we need to
estimate the randomness deviation for arbitrary hidden variables.
The practical weak randomness model is given by
\begin{equation}
\begin{array}{lll}
|p(x_0|\lambda_{0}=i)-\dfrac{1}{2}|\leq\varepsilon_0,\\
|p(x_1|\lambda_{1}=j)-\dfrac{1}{2}|\leq\varepsilon_1,
\end{array}
\end{equation}
where $0\leq \varepsilon_0, \varepsilon_1 \leq\dfrac{1}{2}$,
$\varepsilon_0=0$ ($\varepsilon_1=0$) is the perfect random number
case, which means that Eve has no prior knowledge about the
classical bit selection (bases selection). While
$\varepsilon_0=\dfrac{1}{2}$ ($\varepsilon_1=\dfrac{1}{2}$) means
Eve previously knows the classical bit selection (bases selection),
in which case Alice and Bob can't generate any secret key even if
they can observe $p(x_0)=\dfrac{1}{2}$ ($p(x_1)=\dfrac{1}{2}$).

\textbf{One-step post processing method - } By considering the given
hidden variable $\lambda_{0}=i$, we apply the EDP technology to
illustrate the practical state preparation as the following
equation,

\begin{equation}
\begin{array}{lll}
|\varphi\rangle_{\lambda_{0}=i}=\sqrt{p(x_0=0|\lambda_{0}=i)}|00\rangle+\sqrt{p(x_0=1|\lambda_{0}=i)}|11\rangle,
\end{array}
\end{equation}
where Alice encoding the classical bit $0$ with probability
$p(x_0=0|\lambda_{0}=i)$, and encoding the classical bit $1$ with
probability $p(x_0=1|\lambda_{0}=i)=1-p(x_0=0|\lambda_{0}=i)$. By
considering the given hidden variable $\lambda_{1}=j$, Alice
prepares the quantum state in the rectilinear basis with probability
$p(x_1=0|\lambda_{1}=j)$, and prepares the quantum state in the
diagonal basis with probability
$p(x_1=1|\lambda_{1}=j)=1-p(x_1=0|\lambda_{1}=j)$, thus the final
quantum state preparation under the Pauli quantum channel is

\begin{equation}
\begin{array}{lll}
\rho_{AB_{ij}}=\sum_{u,v}q_{u,v} \Big\{
p(x_1=0|\lambda_{1}=j)I\otimes
X^uZ^v|\varphi\rangle\langle\varphi|_{\lambda_{0}=i}Z^vX^u\otimes
I+\\
~~~~~~~~~~~~~~~~~~~~~~~~~~~~~~~~p(x_1=1|\lambda_{1}=j)I\otimes
HX^uZ^vH|\varphi\rangle\langle\varphi|_{\lambda_{0}=i}HZ^vX^uH
\otimes I\Big\},
\end{array}
\end{equation}
where $u,v\in\{0,1\}$, $H=\dfrac{1}{\sqrt{2}}\begin{pmatrix} 1 & 1 \\
1 & -1\end{pmatrix}$ is the Hadmard matrix,  $\sum_{u,v}q_{u,v}=1$,
$q_{0,0}$ is the probability that Eve applies identity operation
$I=\begin{pmatrix} 1 & 0 \\
0 & 1\end{pmatrix}$, $q_{0,1}$ is the probability that Eve applies
phase error operation $Z=\begin{pmatrix} 1 & 0 \\
0 & -1\end{pmatrix}$, $q_{1,0}$ is the probability that Eve applies
bit error operation $X=\begin{pmatrix} 0 & 1 \\
1 & 0\end{pmatrix}$, $q_{1,1}$ is the probability that Eve applies
bit phase error operation $XZ$. Since Alice's state preparation is
restricted in the two dimensional Hilbert space, we can prove the
final secret key rate under the Pauli quantum channel. Thus, the
quantum bit error rate and phase error rate introduced by Eve can be
respectively given by

\begin{equation}
\begin{array}{lll}
e_{bit}^{i,j}~~~=\langle\phi_2|\rho_{AB_{ij}}|\phi_2\rangle+\langle\phi_4|\rho_{AB_{ij}}|\phi_4\rangle,\\
e_{phase}^{i,j}=\langle\phi_3|\rho_{AB_{ij}}|\phi_3\rangle+\langle\phi_4|\rho_{AB_{ij}}|\phi_4\rangle,
\end{array}
\end{equation}
where
\begin{equation}
\begin{array}{lll}
|\phi_1\rangle=\dfrac{1}{\sqrt{2}}\big(|00\rangle+|11\rangle\big),\\|\phi_2\rangle=\dfrac{1}{\sqrt{2}}\big(|01\rangle+|10\rangle\big),\\
|\phi_3\rangle=\dfrac{1}{\sqrt{2}}\big(|00\rangle-|11\rangle\big),\\|\phi_4\rangle=\dfrac{1}{\sqrt{2}}\big(|01\rangle-|10\rangle\big).
\end{array}
\end{equation}
For arbitrary hidden variable $\lambda_{0}=i$ and $\lambda_{1}=j$,
upper bound of the
 phase error rate $e_{phase}^{i,j}$ can be estimated by applying the bit
 error rate $e_{bit}^{i,j}$ and the randomness deviation parameters,

\begin{equation}
\begin{array}{lll}
e_{phase}^{i,j}-e_{bit}^{i,j}\\=
\Big(\dfrac{1}{2}-\sqrt{-(p(x_0=0|\lambda_{0}=i)-\dfrac{1}{2})^2+\dfrac{1}{4}}\Big)q_{00}\\
~~-\Big(\dfrac{1}{2}-\sqrt{-(p(x_0=0|\lambda_{0}=i)-\dfrac{1}{2})^2+\dfrac{1}{4}}\Big)q_{11}\\
~~+\Big(2p(x_1=0|\lambda_{1}=j)-1\Big)\Big(\dfrac{1}{2}+\sqrt{-(p(x_0=0|\lambda_{0}=i)-\dfrac{1}{2})^2+\dfrac{1}{4}}\Big)q_{01}\\
~~-\Big(2p(x_1=0|\lambda_{1}=j)-1\Big)\Big(\dfrac{1}{2}+\sqrt{-(p(x_0=0|\lambda_{0}=i)-\dfrac{1}{2})^2+\dfrac{1}{4}}\Big)q_{10}\\
\leq
\Big(\dfrac{1}{2}-\sqrt{-\epsilon_0^2+\dfrac{1}{4}}\Big)q_{00}+\Big(\dfrac{1}{2}-\sqrt{-\epsilon_0^2+\dfrac{1}{4}}\Big)q_{11}+2\epsilon_1q_{01}+2\epsilon_1q_{10}\\
\leq
max\Big(\Big(\dfrac{1}{2}-\sqrt{-\epsilon_0^2+\dfrac{1}{4}}\Big),2\epsilon_1\Big)\\
\equiv\delta,
\end{array}
\end{equation}
where we apply $q_{00}+q_{11}\leq1$, $q_{01}+q_{10}\leq1$ and
$\sum_{u,v}q_{u,v}=1$ in the previous calculation. By applying the
EDP technology, the final secret key rate with given hidden
variables $\lambda_{0}=i$ and $\lambda_{1}=j$ is

\begin{equation}
\begin{array}{lll}
R^{i,j}\geq1-h(e_{phase}^{i,j})-h(e_{bit}^{i,j})\geq1-h(e_{bit}^{i,j}+\delta)-h(e_{bit}^{i,j}).
\end{array}
\end{equation}

In the practical experimental realization, we can only observe the
practical quantum bit error rate
$e_{bit}=\sum_{i,j}p_{\lambda_{0}=i}p_{\lambda_{1}=j}e_{bit}^{i,j}$,
 the final secret key rate with given quantum bit error rate $e_{bit}$
can be given by

\begin{equation}
\begin{array}{lll}
R\geq \sum_{i,j}p_{\lambda_{0}=i}p_{\lambda_{1}=j}R^{i,j}\\
~~\geq\sum_{i,j} p_{\lambda_{0}=i}p_{\lambda_{1}=j} \Big(1-h(e_{phase}^{i,j})-h(e_{bit}^{i,j})\Big)\\
~~\geq\sum_{i,j} p_{\lambda_{0}=i}p_{\lambda_{1}=j} \Big(1-h(e_{bit}^{i,j}+\delta)-h(e_{bit}^{i,j})\Big)\\
~~\geq
1-h\Big(\sum_{i,j}p_{\lambda_{0}=i}p_{\lambda_{1}=j}e_{bit}^{i,j}+\delta\Big)-h\Big(\sum_{i,j}p_{\lambda_{0}=i}p_{\lambda_{1}=j}e_{bit}^{i,j}\Big)\\
~~=1-h(e_{bit}+\delta)-h(e_{bit}),
\end{array}
\end{equation}
where we apply the concavity property of the Shannon entropy
function in the previous calculation. By implementing the security
analysis result, we calculate the secret key rate $R$ with given
randomness deviation parameters $\epsilon_0$ and $\epsilon_1$ in
Fig. 2. The calculation result demonstrates that the bases selection
weak randomness decrease the final secret key rate more obviously
comparing with the classical bit encoding weak randomness.

\begin{figure}[!h]\center
\resizebox{11cm}{!}{
\includegraphics{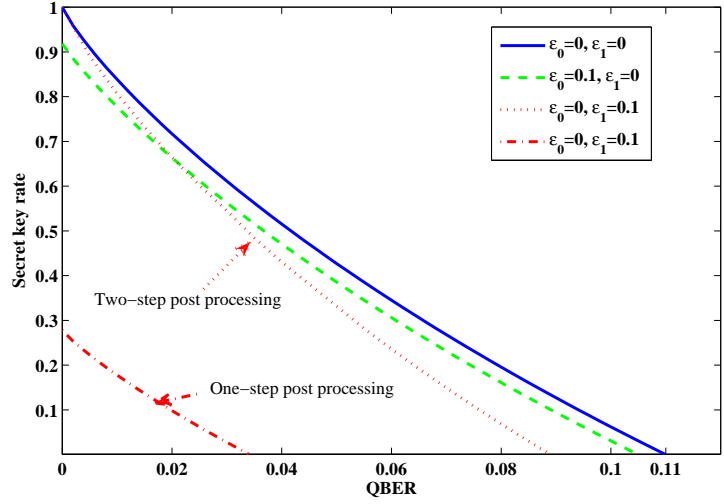}}
\caption{Secret key rate with different quantum bit error rate
value, where the blue solid line is no randomness deviation case,
the green dash line is considering
 $\epsilon_0=0.1$ and $\epsilon_1=0$, the red dotted line is considering $\epsilon_0=0$ and $\epsilon_1=0.1$ with two-step post processing
method, the red dash dotted line is considering $\epsilon_0=0$ and
$\epsilon_1=0.1$ with one-step post processing method.  Comparing
with the one-step post processing method, two-step post processing
method can generate much more secret key with given basis selection
randomness deviation, this is because we can get more precious phase
error estimation in the two-step post processing method.  }
\end{figure}

\textbf{Two-step post processing method -} In the previous weak
randomness model, the input random numbers maybe controlled by the
hidden variables $\lambda_{0}$ and $\lambda_{1}$. Since there are
two different bases selection (diagonal basis and rectilinear basis)
and two different classical bit encoding (0 and 1), we can simply
assume $\lambda_{0}$ and $\lambda_{1}$ have two different values $
\{0,1\}$ respectively.

 In the practical
experimental realization, we can only observe the classical bit
encoding probability
$p(x_0)=p_{\lambda_{0}=0}p(x_0|\lambda_{0}=0)+p_{\lambda_{0}=1}p(x_0|\lambda_{0}=1)$,
but $p(x_0)=\dfrac{1}{2}$ can't guarantee
$p(x_0|\lambda_{0}=0)=p(x_0|\lambda_{0}=1)=\dfrac{1}{2}$, the
detailed classical bit deviation model is given in Fig. 3.
\begin{figure}[!h]\center
\resizebox{9cm}{!}{
\includegraphics{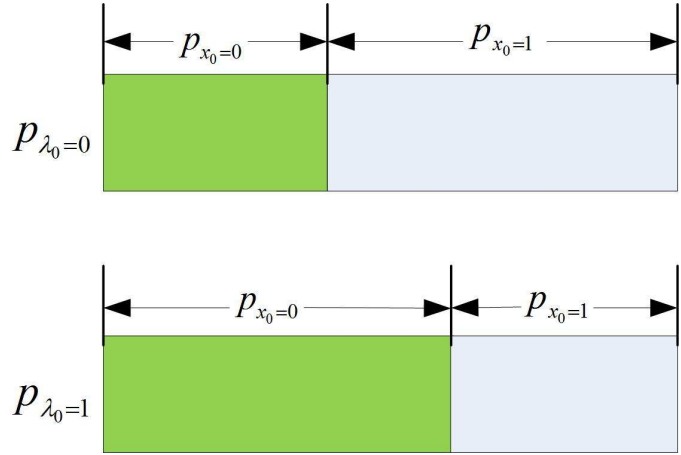}}
\caption{The classical bit encoding $x_0$ is controlled by the
hidden variable $\lambda_{0}$, different $\lambda_{0}$ values have
different classical bit encoding probability $p(x_0|\lambda_{0})$. }
\end{figure}
Similarly, we can also only observe the bases selection probability
$p(x_1)=p_{\lambda_{1}=0}p(x_1|\lambda_{1}=0)+p_{\lambda_{1}=1}p(x_1|\lambda_{1}=1)$,
but the observed probability $p(x_1)=\dfrac{1}{2}$ can't guarantee
$p(x_1|\lambda_{1}=0)=p(x_1|\lambda_{1}=1)=\dfrac{1}{2}$, the
detailed bases selection deviation model is given in Fig. 4.
\begin{figure}[!h]\center
\resizebox{9cm}{!}{
\includegraphics{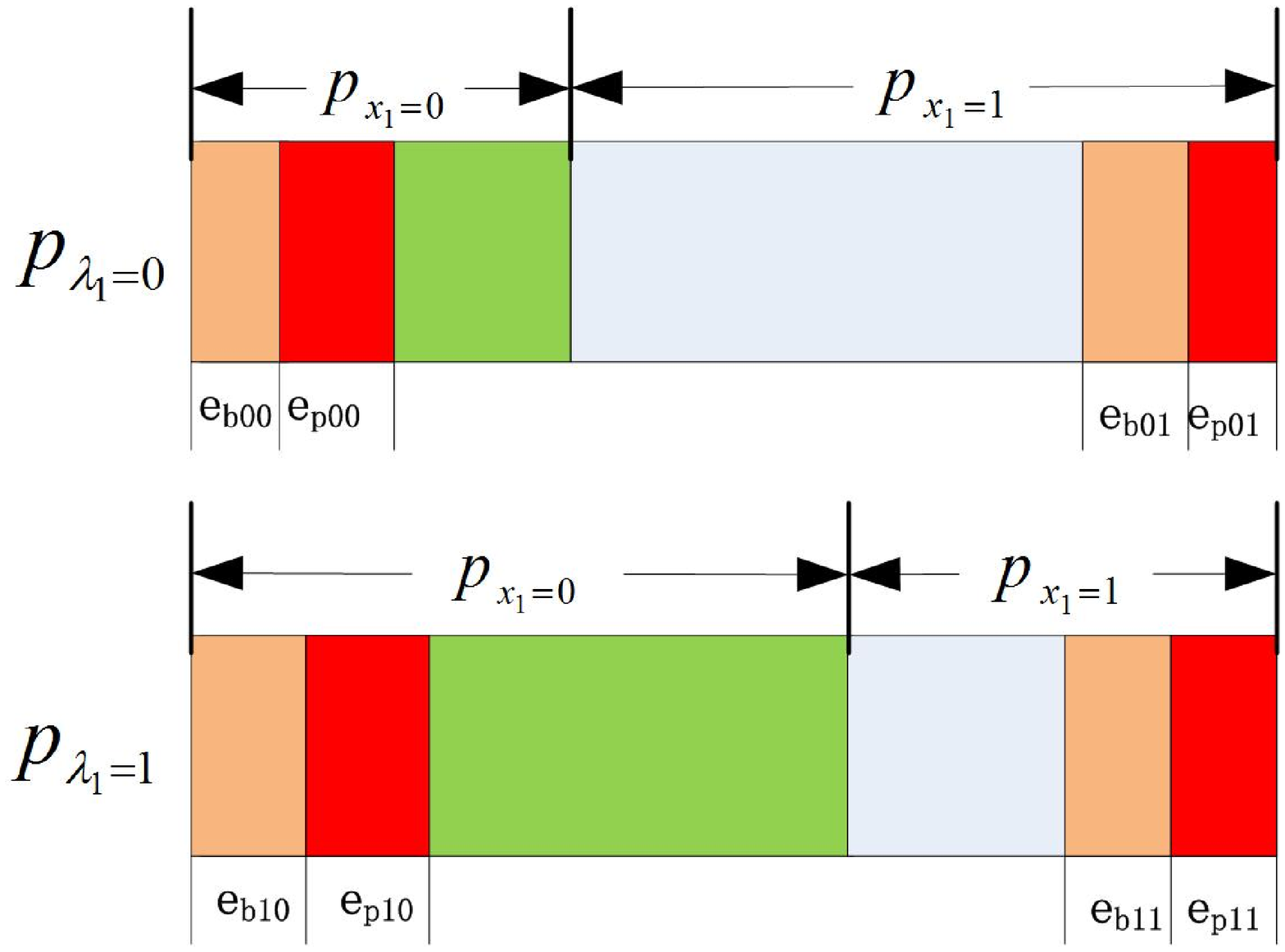}}
\caption{The basis selection deviation is controlled by the hidden
variable $\lambda_{1}$, different $\lambda_{1}$ value has different
basis selection probability $p(x_1|\lambda_{1})$. For given hidden
variable $\lambda_1=0$, $e_{b00}$ and $e_{b01}$ are bit error rates
introduced in the rectilinear basis and diagonal basis, while
$e_{p00}$ and $e_{p01}$ are phase error rates introduced in the
rectilinear basis and diagonal basis respectively. For given hidden
variable $\lambda_1=1$, $e_{b10}$ and $e_{b11}$ are bit error rates
introduced in the rectilinear basis and diagonal basis, while
$e_{p10}$ and $e_{p11}$ are phase error rates introduced in the
rectilinear basis and diagonal basis respectively.}
\end{figure}
The practical quantum state preparation is given by

\begin{equation}
\begin{array}{lll}
\rho_{AB}=\sum_{{\lambda_{1}}}p_{\lambda_{1}}p(x_1=0|\lambda_{1})\sum_{{\lambda_{0}}}p_{\lambda_{0}}\rho_{AB_{Z\lambda_{0}}}+
\sum_{{\lambda_{1}}}p_{\lambda_{1}}p(x_1=1|\lambda_{1})\sum_{{\lambda_{0}}}p_{\lambda_{0}}\rho_{AB_{X\lambda_{0}}},\\
\end{array}
\end{equation}
where
\begin{equation}
\begin{array}{lll}
\rho_{AB_{Z\lambda_{0}}}=\sum_{u,v}q_{u,v} I\otimes
X^uZ^v|\varphi\rangle\langle\varphi|_{\lambda_{0}}Z^vX^u\otimes I,\\
\rho_{AB_{X\lambda_{0}}}=\sum_{u,v}q_{u,v} I\otimes
HX^uZ^vH|\varphi\rangle\langle\varphi|_{\lambda_{0}}HZ^vX^uH\otimes
I.
\end{array}
\end{equation}
For given hidden variables $\lambda_{0}$ and $\lambda_{1}$, the
difference between the phase error rate in the rectilinear basis and
bit error rate in the diagonal basis can be given by
\begin{equation}
\begin{array}{lll}
|e_{p{\lambda_{0}}{\lambda_{1}}0}-e_{b{\lambda_{0}}{\lambda_{1}}1}|\leq
\dfrac{1}{2}-\sqrt{-\epsilon_0^2+\dfrac{1}{4}}.
\end{array}
\end{equation}
where
$e_{p{\lambda_{0}}{\lambda_{1}}0}=\langle\phi_3|\rho_{AB_{Z\lambda_{0}}}|\phi_3\rangle+\langle\phi_4|\rho_{AB_{Z\lambda_{0}}}|\phi_4\rangle$,
$e_{b{\lambda_{0}}{\lambda_{1}}1}=\langle\phi_2|\rho_{AB_{X\lambda_{0}}}|\phi_2\rangle+\langle\phi_4|\rho_{AB_{X\lambda_{0}}}|\phi_4\rangle$.
 Similarly, The difference between the phase error rate in the
diagonal basis and bit error rate in the rectilinear basis can be
given by
\begin{equation}
\begin{array}{lll}
|e_{p{\lambda_{0}}{\lambda_{1}}1}-e_{b{\lambda_{0}}{\lambda_{1}}0}|
\leq \dfrac{1}{2}-\sqrt{-\epsilon_0^2+\dfrac{1}{4}},
\end{array}
\end{equation}
where
$e_{p{\lambda_{0}}{\lambda_{1}}1}=\langle\phi_3|\rho_{AB_{X\lambda_{0}}}|\phi_3\rangle+\langle\phi_4|\rho_{AB_{X\lambda_{0}}}|\phi_4\rangle$,
$e_{b{\lambda_{0}}{\lambda_{1}}0}=\langle\phi_2|\rho_{AB_{Z\lambda_{0}}}|\phi_2\rangle+\langle\phi_4|\rho_{AB_{Z\lambda_{0}}}|\phi_4\rangle$.
By considering
$e_{p{\lambda_{1}}0}=\langle\phi_3|\sum_{{\lambda_{0}}}p_{\lambda_{0}}\rho_{AB_{Z\lambda_{0}}}|\phi_3\rangle+\langle\phi_4|\sum_{{\lambda_{0}}}p_{\lambda_{0}}\rho_{AB_{Z\lambda_{0}}}|\phi_4\rangle=\sum_{\lambda_{0}}p_{\lambda_{0}}e_{p{\lambda_{0}}{\lambda_{1}}0}$
and
$e_{b{\lambda_{1}}1}=\langle\phi_2|\sum_{{\lambda_{0}}}p_{\lambda_{0}}\rho_{AB_{X\lambda_{0}}}|\phi_2\rangle+\langle\phi_4|\sum_{{\lambda_{0}}}p_{\lambda_{0}}\rho_{AB_{X\lambda_{0}}}|\phi_4\rangle=\sum_{\lambda_{0}}p_{\lambda_{0}}e_{b{\lambda_{0}}{\lambda_{1}}1}$,
we calculate the difference between the phase error rate
$e_{p{\lambda_{1}}0}$ and the bit error rate $e_{b{\lambda_{1}}1}$
\begin{equation}
\begin{array}{lll}
|e_{p{\lambda_{1}}0}-e_{b{\lambda_{1}}1}|=\big | \sum_{\lambda_{0}}p_{\lambda_{0}}\big(e_{p{\lambda_{0}}{\lambda_{1}}0}-e_{b{\lambda_{0}}{\lambda_{1}}1}\big)\big|\\
~~~~~~~~~~~~~~~~~~~~~~\leq\sum_{\lambda_{0}}p_{\lambda_{0}}\big|e_{p{\lambda_{0}}{\lambda_{1}}0}-e_{b{\lambda_{0}}{\lambda_{1}}1}\big|\\
~~~~~~~~~~~~~~~~~~~~~~\leq\sum_{\lambda_{0}}p_{\lambda_{0}}\big(\dfrac{1}{2}-\sqrt{-\epsilon_0^2+\dfrac{1}{4}}\big)\\
~~~~~~~~~~~~~~~~~~~~~~=
\dfrac{1}{2}-\sqrt{-\epsilon_0^2+\dfrac{1}{4}}.
\end{array}
\end{equation}
Similarly, the difference between $e_{p{\lambda_{1}}1}$ and
$e_{b{\lambda_{1}}0}$ is
\begin{equation}
\begin{array}{lll}
|e_{p{\lambda_{1}}1}-e_{b{\lambda_{1}}0}| \leq
\dfrac{1}{2}-\sqrt{-\epsilon_0^2+\dfrac{1}{4}}.
\end{array}
\end{equation}

The probability of getting the rectilinear basis and diagonal basis
measurement outcomes in Bob's side can be respectively given by
\begin{equation}
\begin{array}{lll}
p_{rec}=p_{rec1}+p_{rec2},~~p_{dia}=p_{dia1}+p_{dia2},
\end{array}
\end{equation}
where
$p_{rec1}=p_{{\lambda_1}=0}p(x_1=0|{\lambda_1}=0),~p_{rec2}=p_{{\lambda_1}=1}p(x_1=0|{\lambda_1}=1),
~~p_{dia1}=p_{{\lambda_1}=0}p(x_1=1|{\lambda_1}=0),~p_{dia2}=p_{{\lambda_1}=1}p(x_1=1|{\lambda_1}=1)$.
The phase error rate in the rectilinear basis and diagonal basis can
be respectively given by
\begin{equation}
\begin{array}{lll}
e_{recpha}=\dfrac{p_{rec1}e_{p00}+p_{rec2}e_{p10}}
{p_{rec}}\leq\dfrac{p_{rec1}e_{b01}+p_{rec2}e_{b11}}
{p_{rec}}+\dfrac{1}{2}-\sqrt{-\epsilon_0^2+\dfrac{1}{4}},\\
e_{diapha}=\dfrac{p_{dia1}e_{p01}+p_{dia2}e_{p11}}
{p_{dia}}\leq\dfrac{p_{dia1}e_{b00}+p_{dia2}e_{b10}}
{p_{dia}}+\dfrac{1}{2}-\sqrt{-\epsilon_0^2+\dfrac{1}{4}}.
\end{array}
\end{equation}
The bit error rate in the rectilinear basis and diagonal basis can
be respectively given by
\begin{equation}
\begin{array}{lll}
e_{recbit}=\dfrac{p_{rec1}e_{b00}+p_{rec2}e_{b10}}
{p_{rec}},e_{diabit}=\dfrac{p_{dia1}e_{b01}+p_{dia2}e_{b11}}
{p_{dia}}.
\end{array}
\end{equation}
By applying the two-step post processing method with the two
different bases measurement outcomes, the final secret key rate can
be given by
\begin{equation}
\begin{array}{lll}
R\geq
p_{rec}\big(1-h(e_{recbit})-h(e_{recpha})\big)+p_{dia}\big(1-h(e_{diabit})-h(e_{diapha})\big),
\end{array}
\end{equation}
where the first part is the secret key generated by the rectilinear
basis, while the second part is the secret key generated by the
diagonal basis. The corresponding secret key rate $R$ with different
quantum bit error rate values is given in Fig. 2, the calculation is
based on the nonlinear optimization method with given quantum bit
error rate, the detailed explanation is in the methods. To explain
our analysis result, we compare the two analysis methods by
considering the wavelength attack has the coupling ratio $0.4$ and
$0.6$ with different wavelengths. If the observed quantum bit error
rate is $0.02$, one-step post processing method can generate the
secret key rate
 $0.0984$, while the two-step post processing method can generate the secret key
 rate $0.6642$.

\textbf{Methods-} By considering Eve's arbitrary attacking scheme,
the final secret key rate with two different bases can be calculated
with the following optimization method
\begin{equation}
\begin{array}{lll}
Minimize:~~p_{rec}\big(1-h(e_{recbit})-h(e_{recpha})\big)+p_{dia}\big(1-h(e_{diabit})-h(e_{diapha})\big)\\
Subject~~to:~p_{{\lambda_0}=0}+p_{{\lambda_0}=1}=p_{{\lambda_1}=0}+p_{{\lambda_1}=1}=1\\
~~~~~~~~~~~~~~~~~~~~~~p(x_0=0|{\lambda_0})+p(x_0=1|{\lambda_0})=p(x_1=0|{\lambda_1})+p(x_1=1|{\lambda_1})=1\\
~~~~~~~~~~~~~~~~~~~~~~0\leq
e_{b00},e_{b01},e_{b10},e_{b11},p_{{\lambda_0}=0},p_{{\lambda_1}=0}\leq
1\\
~~~~~~~~~~~~~~~~~~~~~~|p(x_0|\lambda_{0}=i)-\dfrac{1}{2}|\leq\varepsilon_0\\
~~~~~~~~~~~~~~~~~~~~~~|p(x_1|\lambda_{1}=j)-\dfrac{1}{2}|\leq\varepsilon_1\\
~~~~~~~~~~~~~~~~~~~~~~|e_{p{\lambda_{1}}0}-e_{b{\lambda_{1}}1}| \leq
\dfrac{1}{2}-\sqrt{-\epsilon_0^2+\dfrac{1}{4}}\\
~~~~~~~~~~~~~~~~~~~~~~|e_{p{\lambda_{1}}1}-e_{b{\lambda_{1}}0}| \leq
\dfrac{1}{2}-\sqrt{-\epsilon_0^2+\dfrac{1}{4}}\\
~~~~~~~~~~~~~~~~~~~~~~p_{rec}=p_{dia}=\dfrac{1}{2}\\
~~~~~~~~~~~~~~~~~~~~~~p_{rec}e_{recbit}+ p_{dia}e_{diabit}=Q,\\
\end{array}
\end{equation}
where $Q$ is the quantum bit error rate estimated in the practical
experimental realization, $p_{rec}=p_{dia}=\dfrac{1}{2}$ are the
bases selection probability observed in the practical experimental
realization.

\textbf{Conclusion -} In this work, security of BB84 QKD protocol
again the strong randomness attack and the weak randomness attack
have been analyzed, which satisfies several practical attacking
schemes, such as the photon number splitting attack, detector
blinding attack, wavelength attack and time shift attack. We
demonstrate that security of the practical QKD system can be
evaluated by respectively considering the Trojan horse attack, the
strong randomness attack and the weak randomness attack, and the
three attacking models can be employed to build the practical QKD
system security standardization in the future.

 \textbf{Acknowledgements - } The authors are
supported by the the National Natural Science Foundation of China
(Grant Nos. 11304397, 61201239, 61205118 and 61475148), China
Postdoctoral Science Foundation (Grant No. 2013M540514) and Anhui
Provincial Natural Science Foundation(Grant No. 1408085QF102).


\begin{thebibliography}{10}

\bibitem{reviewBB84} Bennett, C. H., Brassard, G., Quantum cryptography: Public-key distribution and coin tossing,
  Proceedings IEEE Int. Conf. on Computers, Systems and Signal Processing, Bangalore, India, pp. 175-179 (IEEE, New York, 1984).

\bibitem{edp1} Lo, H. K., Chau, H. F., Unconditional security of quantum key distribution over arbitrarily long distances. Science 283, 5410 (1999).


\bibitem{edp2} Shor, P., Preskill, J., Simple proof of security of the BB84 quantum key distribution protocol. Phys. Rev. Lett. 85, 441-444 (2000).

\bibitem{re1} Renner, R., Security of Quantum Key Distribution. PhD thesis,
Diss.~ETH No 16242, quant-ph/0512258 (2005).

\bibitem{re2} Kraus, B., Gisin, N., Renner, R., Lower and Upper Bounds on the
Secret-Key Rate for Quantum Key Distribution Protocols Using One-Way
Classical Communication. Phys. Rev. Lett. 95, 080501(2005).

\bibitem{re3}Renner, R., Gisin, N., Kraus, B., Information-theoretic security proof
for quantum-key-distribution protocols. Phys. Rev. A 72, 012332
(2005).

\bibitem{practical}
Scarani, V. et al. The security of practical quantum key
distribution. Rev. Mod. Phys. 81, 1301-1350 (2009).


\bibitem{trojan} Gisin,  N., Fasel, S., Kraus, B., Zbinden, H., and Ribordy, G., Trojan-horse attacks on quantum-key-distribution systems. Phys. Rev. A, 73, 022320 (2006).

\bibitem{PNS1} Lutkenhaus, N., Estimates for practical quantum cryptography. Phys. Rev. A, 59, 3301 (1999).

\bibitem{PNS2} Lutkenhaus, N., Security against individual attacks for realistic
quantum key distribution. Phys. Rev. A, 61, 052304 (2000).

\bibitem{blind1} Lydersen, L., Wiechers, C., Wittmann, C., Elser, D., Skaar, J., and
Makarov, V., Hacking commercial quantum cryptography systems by
tailored bright illumination.  Nature Photonics, 4, 686-689 (2010).

\bibitem{blind2} Yuan, Z. L., Dynes J. F., and Shields, A. J., Avoiding the blinding
attack in QKD. Nature Photonics, 4, 800 (2010).

\bibitem{probability} Qian, Y. J., Li, H. W., He, D. Y., Yin, Z. Q., Zhang, C. M., Chen, W., Wang, S., Han, Z. F., Countermeasure against probabilistic blinding attack in practical
quantum key distribution systems, Chin. Phys. B, Vol. 24, No. 9,
090305 (2015).


\bibitem{gllp} Gottesman, D., Lo, H. K., Lutkenhaus, N., and Preskill, J., Security of
quantum key distribution with imperfect devices. Quant. Inf.
Comput., 5, 325-360 (2004).



\bibitem{decoy1} Hwang, W. Y., Quantum Key Distribution with High Loss: Toward Global Secure Communication. Phys. Rev. Lett., 91, 057901 (2003).

\bibitem{decoy2} Wang, X. B., Phys. Rev. Lett., Beating the photon-number-splitting attack in practical quantum cryptography. 94, 230503 (2005).

\bibitem{decoy3} Lo, H. K., Ma, X., and Chen, K., Decoy State Quantum Key Distribution. Phys. Rev. Lett., 94, 230504 (2005).


\bibitem{winter} Devetak, I., and Winter, A., Distillation of secret key and entanglement from quantum
states. Proc. R. Soc. Lond. A 461 (2005).

\bibitem{weak} Bouda, J., Pivoluska, M., Plesch, M. and Wilmott. C,
Weak randomness seriously limits the security of quantum key
distribution. Phys. Rev. A, 86, 062308 (2012).

\bibitem{wavelength} Li, H. W., et al., Attacking a practical quantum-key-distribution
system with wavelength-dependent beam-splitter and multiwavelength
sources. Phys. Rev. A, 84, 062308 (2011).

\bibitem{time} Qi, B.,  Fung, C. H. F., Lo, H. K., and Ma, X.,  Time-shift attack in
practical quantum cryptosystems.  Quantum Inf. Comput., 7, 73-82
(2007).

\end{thebibliography}
\end{document}